\documentclass[letterpaper, 10 pt, conference]{ieeeconf}  

\IEEEoverridecommandlockouts                              

\usepackage{amsmath} 
\usepackage{amssymb}  
\usepackage{mathrsfs}
\usepackage{mathtools}
\usepackage{mathdots}
\usepackage{bm}
\usepackage{url}
\usepackage{graphicx}
\usepackage{xcolor} 
\usepackage{cite}

\usepackage{pgfplots}
\pgfplotsset{compat=1.18}
\usepgfplotslibrary{groupplots}

\DeclareRobustCommand{\frfSampleMarker}{%
\tikz[baseline=-0.5ex,x=1em,y=1em]{%
    \draw[black,line width=0.6pt]
    (-0.16,-0.16)--(0.16,0.16)
    (-0.16,0.16)--(0.16,-0.16);}}
\DeclareRobustCommand{\frfVarianceLine}{%
\tikz[baseline=-0.5ex,x=1em,y=1em]{%
    \draw[red,line width=0.6pt,dash pattern=on 3pt off 2pt]
    (0,0)--(1.5,0);}}

\definecolor{predictionRed}{rgb}{0.80,0.15,0.15}
\DeclareRobustCommand{\predictionLegend}[2]{%
\tikz[baseline=-0.5ex,x=1em,y=1em]{%
    \filldraw[fill=#2,draw=#1,line width=0.3pt]
    (0,-0.18) rectangle (1.5,0.18);
    \draw[#1,line width=0.6pt] (0,0)--(1.5,0);
    \fill[#1] (0.75,0) circle[radius=0.07em];}}

\DeclareRobustCommand{\controlBox}[1]{%
\tikz[baseline=-0.5ex,x=1em,y=1em]{%
    \draw[#1,line width=0.5pt] (0,0)--(1.5,0)
    (0,-0.1)--(0,0.1) (1.5,-0.1)--(1.5,0.1);
    \filldraw[fill=#1!25,draw=#1,line width=0.5pt]
    (0.35,-0.18) rectangle (1.15,0.18);
    \draw[#1,line width=0.7pt] (0.75,-0.18)--(0.75,0.18);}}
\DeclareRobustCommand{\controlOracleLine}{%
\tikz[baseline=-0.5ex,x=1em,y=1em]{%
    \draw[black,line width=0.6pt,dash pattern=on 3pt off 2pt]
    (0,0)--(1.5,0);}}
\definecolor{controlGray}{rgb}{0.45,0.45,0.45}

\newtheorem{proposition}{\bf Proposition}
\newtheorem{lemma}{\bf Lemma}

\newtheorem{definition}{\bf Definition}
\newtheorem{remark}{\bf Remark}

\DeclareMathOperator{\im}{im}
\DeclareMathOperator{\rank}{rank}

\title{\LARGE \bf
Sample-Covariance Regularization for Data-Driven Predictive Control from Frequency Response Functions under Colored Noise
}

\author{M. Wind$^{1}$, T.J. Meijer$^{2}$, V.S. Dolk$^{1,3}$, W.P.M.H. Heemels$^{1}$ 
\thanks{$^{1}$Eindhoven University of Technology, Control Systems Technology Section,
    {\tt\small m.wind@tue.nl}}%
\thanks{$^{2}$Lund University, Department of Automatic Control, Sweden. T.J. Meijer is a member of ELLIIT Strategic Research Area, and received funding from ERC Advanced Grant agreement No 101199738 (DualControl).}%
\thanks{$^{3}$ASML Veldhoven, The Netherlands}%
\thanks{This research was conducted under project 24PPS052 COMEDI, “Cost-effective Mechatronic Design for High-Tech Systems.”}
}

\usepackage{eso-pic}
\AddToShipoutPictureBG*{%
\AtPageUpperLeft{%
\setlength\unitlength{1in}%
\hspace*{\dimexpr0.5\paperwidth\relax}
\makebox(0,-1)[c]{
\begin{tabular}{c c}
M.~Wind \emph{et al.},
Sample-Covariance Regularization for Data-Driven Predictive Control \\ from Frequency Response Functions under Colored Noise, Preprint uploaded to ArXiv 6 November 2026 \\
\end{tabular}}}}

\begin{document}
\setlength{\parskip}{0pt}
\setlength{\parindent}{1em}
\maketitle
\thispagestyle{empty}
\pagestyle{empty}
\begin{abstract}
In this work, we investigate how sample covariance obtained from frequency-response functions (FRFs) can be exploited to improve performance of data-driven frequency-domain predictive control (FreePC) in the presence of colored measurement noise. In case of periodic multisine excitation, this covariance is obtained directly from repeated periods and provides a frequency-dependent measure of uncertainty without requiring an additional parametric noise model. To incorporate this information into both indirect and
direct predictive control, we develop a frequency-domain subspace predictive controller (FD-SPC) and establish its equivalence to the direct FreePC formulation in the noiseless setting. Regularization terms for both formulations are then motivated by expected prediction-error criteria and constructed from the sample covariance of the FRF estimates. Across all tested noise models and data sizes, regularization consistently reduces the median realized control cost relative to the corresponding controller without covariance regularization, with reductions of up to 35\%.
\end{abstract}

\section{INTRODUCTION}
Frequency-response functions (FRFs) are widely used in industrial control practice because they can be estimated directly from standard excitation experiments and provide an interpretable description of the system over the frequency range of interest. Furthermore, under periodic excitation, e.g., multisine excitation signals, the FRF estimate is asymptotically unbiased and a frequency-dependent estimate of its uncertainty, i.e., a nonparametric noise characteristic, can directly be obtained, see \cite{pintelonSystemIdentificationFrequency2012, pintelonEstimationNonparametricNoise2010a,schoukensFrequencydomainSystemIdentification1997, caubergheFrequencydomainSubspaceIdentification2006}.

Despite these advantages, using FRFs in constrained predictive control is not immediate. Conventional model predictive control (MPC) is typically formulated using a parametric state-space model, together with a state estimator when the state is not directly measured. Utilizing FRF information would typically involve first identifying a state-space model, for instance via frequency-domain subspace identification procedures, see, e.g., \cite{mckelveySubspacebasedMultivariableSystem1996a, vanoverscheeContinuoustimeFrequencyDomain1996,mckelveyFrequencyDomainSystem1997,mckelveySubspaceMethodsFrequency2004}. The FRF sample covariance can then be used to weight the
identification criterion \cite{pintelonFrequencydomainSubspaceSystem2002}. 

Alternatively, a predictive representation can be constructed directly from measured input-output data, without first identifying a parametric state-space model. Willems' fundamental lemma provides a data-based representation of finite-length system trajectories without identifying a parametric model \cite{willemsNotePersistencyExcitation2005}. This representation can be used directly in data-enabled
predictive control (DeePC) \cite{coulsonDataEnabledPredictiveControl2019}, or indirectly by first estimating a multistep predictor, as in subspace predictive control (SPC) \cite{favoreelSPCSubspacePredictive1999}. For comparisons between DeePC and SPC, see \cite{krishnanDirectVsIndirect2021, dorflerBridgingDirectIndirect2023, fiedlerRelationshipDataenabledPredictive2021,wingerdenDataenabledPredictiveControl2022} and a comprehensive overview of both methods is provided in \cite{verheijenHandbookLinearDatadriven2023}. 

Until recently, data-driven predictive-control formulations were predominantly based on time-domain data. A frequency-domain version of Willems' fundamental lemma (FD-WFL) was introduced in \cite{ferizbegovicWillemsFundamentalLemma2021,meijerFrequencyDomainWillemsLemma2025a} and extended in \cite{meijerLeveragingNonSteadyStateFrequencyDomain2025a} to accommodate non-steady-state data. Frequency-domain data-enabled predictive control (FreePC) \cite{meijerFrequencyDomainWillemsLemma2025a} uses this trajectory representation for constrained predictive control. At each sampling instant, it selects a linear combination of the frequency-domain data columns that matches the measured input-output history and minimizes the tracking cost over the resulting future trajectory. In this work, we derive an indirect counterpart, FD-SPC, that computes a multistep predictor from the same data before solving the online control problem.

Measurement noise introduces uncertainty into the estimated FRFs and the resulting trajectory representation. Sparse column selection reduces the dimension of the FreePC parametrization \cite{uchidaLASSOBasedDataReduction2026}, but does not address uncertainty in the underlying FRF estimates. Regularized and distributionally robust DeePC formulations account for uncertain data more generally \cite{coulsonRegularizedDistributionallyRobust2019,huangRobustDataEnabledPredictive2023,damjanovicDistributionallyRobustFrequency2026}, but do not exploit the frequency-dependent uncertainty information available from repeated FRF measurements.

To address this gap, we exploit the information about sample variance of the estimated FRFs, which is readily available from repeated periods under periodic excitation. This information is used to regularize the construction of the multistep predictor in the indirect FD-SPC formulation and the trajectory parametrization in the direct FreePC formulation. In particular, the contributions of this paper are:
\begin{enumerate}[\setlength{\topsep}{0pt}]
\item An indirect frequency-domain subspace predictive controller (FD-SPC), without identifying a parametric state-space realization.
\item Establishing equivalence between FD-SPC and FreePC in the noiseless setting, connecting the indirect and direct frequency-domain formulations.
\item Exploiting FRF sample covariance by regularizing the FD-SPC predictor fit and the trajectory coefficients in FreePC.
\item A numerical evaluation under white and colored noise, demonstrating the effect of regularization on control performance for different number of measured periods.
\end{enumerate}

\section{PRELIMINARIES AND PROBLEM SETTING}
This section introduces the notation and system description, reviews the time-domain data equation underlying subspace predictive control, and summarizes the frequency-domain Willems' lemma and FRF estimation.
\subsection{Notation}
Let $\mathbb{R}$ denote the field of real numbers, $\mathbb{C}$ the complex plane, and $\mathbb{Z}$ the integers. We define $\mathbb{Z}_{[n,m]} \coloneqq \{n,n+1,\ldots,m\}$ and $\mathbb{Z}_{\geq n} \coloneqq \{n,n+1,\ldots\}$, where $n,m\in\mathbb{Z}$ and $n\leq m$. The imaginary unit is denoted by $j$, i.e., $j^2=-1$. For a complex-valued matrix $A\in\mathbb{C}^{n\times m}$, the symbols $A^\top$, $A^{\mathrm{H}}$, $A^*$, and $A^\dagger$ denote its transpose, complex-conjugate transpose, complex conjugate, and pseudoinverse, respectively, whereas $\operatorname{Re}(A)$ and $\operatorname{Im}(A)$ denote its real and imaginary parts. We use $\otimes$ to denote the Kronecker product and $\|v\|_Q^2=v^\top Qv$ for $Q\succeq 0$. For a sequence $v$, we denote $v_{[a,b]}\coloneqq\operatorname{col}(v_a,v_{a+1},\ldots,v_b)$
for integers $a\leq b$.

\subsection{System description}
Consider the controllable and observable LTI system
\begin{equation}\label{eq:system}
\Sigma:\quad
\begin{cases}
x_{k+1}=Ax_k+Bu_k,\\
y_k=Cx_k+Du_k,
\end{cases}
\end{equation}
where $x_k\in\mathbb{R}^{n_x}$, $u_k\in\mathbb{R}^{n_u}$, and
$y_k\in\mathbb{R}^{n_y}$. Let $\ell_\Sigma \in \mathbb{Z}_{[1,n_x]}$ denote the observability index of $\Sigma$, i.e., the smallest integer for which the observability matrix
\begin{equation}
\mathcal{O}_s \coloneqq\begin{bmatrix}
C^\top & (CA)^\top & \cdots & (CA^{s-1})^\top
\end{bmatrix}^{\top},
\end{equation}
satisfies $\operatorname{rank}\mathcal{O}_{\ell_\Sigma} = \operatorname{rank}\mathcal{O}_{n_x}$.

\subsection{Data equation and subspace predictor}\label{sec:subspace-pred}
For any horizon length $s\in\mathbb Z_{\geq1}$ and initial time $i\in\mathbb Z$, the multistep output equation is
\begin{equation}
y_{[i,i+s-1]}
=
\mathcal O_sx_i+\mathcal T_su_{[i,i+s-1]},
\end{equation}
with $\mathcal{T}_s$ the block-Toeplitz matrix defined as
\begin{equation}
\mathcal T_s \coloneqq
\begin{bmatrix}
    D & 0 & \cdots & 0\\
    CB & D & \cdots & 0\\
    \vdots & \vdots & \ddots & \vdots\\
    CA^{s-2}B & CA^{s-3}B & \cdots & D
\end{bmatrix}.
\end{equation}

To construct a predictor, consider a length-$(p+f)$ trajectory split at time $i$. The past output satisfies
\begin{equation}\label{eq:predat}
y_{[i-p,i-1]}=
\mathcal O_p x_{i-p}
+\mathcal T_pu_{[i-p,i-1]},
\end{equation}
and the state
\begin{equation}\label{eq:statepredat}
x_i=
A^p x_{i-p}
+\mathcal C_pu_{[i-p,i-1]},
\end{equation}
where
\begin{equation}
\mathcal C_p
\coloneqq
\begin{bmatrix}
    A^{p-1}B & A^{p-2}B & \cdots & B
\end{bmatrix}
\end{equation}
is the extended controllability matrix. Let $p\geq \ell_\Sigma$ and since $(A,C)$ is observable, the matrix $\mathcal O_p$ has full column rank and satisfies $\mathcal O_p^\dagger\mathcal O_p=I_{n_x}$. Therefore,
\begin{equation}
x_{i-p}
=
\mathcal O_p^\dagger
\left(
    y_{[i-p,i-1]}
    -\mathcal T_pu_{[i-p,i-1]}
\right).
\end{equation}
Substitution into \eqref{eq:statepredat} gives
\begin{equation}
x_i
=
\mathcal K
\begin{bmatrix}
    u_{[i-p,i-1]}\\
    y_{[i-p,i-1]}
\end{bmatrix},
\qquad
\mathcal K
\coloneqq
\begin{bmatrix}
    \mathcal C_p-A^p\mathcal O_p^\dagger\mathcal T_p &
    A^p\mathcal O_p^\dagger
\end{bmatrix}.
\end{equation}

The corresponding future output satisfies
\begin{equation}\label{eq:io-data-1step}
y_{[i,i+f-1]}
=
\mathcal O_f\mathcal K
\begin{bmatrix}
    u_{[i-p,i-1]}\\
    y_{[i-p,i-1]}
\end{bmatrix}
+\mathcal T_fu_{[i,i+f-1]}.
\end{equation}
Let $u_{[0,N-1]}^{\mathrm d}$ and $y_{[0,N-1]}^{\mathrm d}$ denote a length-$N$ offline (measured) input-output trajectory of $\Sigma$, where $N\geq p+f$. Stacking \eqref{eq:io-data-1step} for all admissible $i$ yields the block-Hankel data equation \cite{verhaegenFilteringSystemIdentification2007}.
\begin{equation}\label{eq:td-data-equation}
Y_{\mathrm f}^{\mathrm d}=\mathcal O_f\mathcal K
\begin{bmatrix}
    U_{\mathrm p}^{\mathrm d}\\
    Y_{\mathrm p}^{\mathrm d}
\end{bmatrix}
+\mathcal T_fU_{\mathrm f}^{\mathrm d}.
\end{equation}
Let $\bar N\coloneqq N-p-f+1$ and define the data matrix as
\begin{equation}
\mathcal D
\coloneqq
\begin{bmatrix}
    U_{\mathrm p}^{\mathrm d}\\
    Y_{\mathrm p}^{\mathrm d}\\
    U_{\mathrm f}^{\mathrm d}
\end{bmatrix}
\in
\mathbb R^{((n_u+n_y)p+n_uf)\times\bar N},
\end{equation}
where
\begin{equation}
Y_{\mathrm f}^{\mathrm d}
\in\mathbb R^{fn_y\times\bar N}.
\end{equation}
Then \eqref{eq:td-data-equation} can be written compactly as
\begin{equation}
Y_{\mathrm f}^{\mathrm d}=\mathcal{P}\mathcal D,
\end{equation}
where the unknown matrix
\begin{equation}\label{eq:P-dimension}
\mathcal{P}
\in
\mathbb R^{fn_y\times((n_u+n_y)p+n_uf)}
\end{equation}
enters linearly. The matrix $\mathcal{P}$ contains the coefficients of a finite-horizon VARX model structure in which future outputs depend linearly on past inputs, past outputs, and future inputs. A predictor can therefore be obtained directly from the least-squares problem
\begin{equation}\label{eq:predictor-ls}
\min_{\mathcal{P}}
\left\|
    Y_{\mathrm f}^{\mathrm d}-\mathcal{P}\mathcal D
\right\|_{\mathrm F}^2,
\end{equation}
where $\left\| \cdot \right\|_F$ denotes the Frobenius norm. The complete set of minimizers of \eqref{eq:predictor-ls} is spanned by 
\begin{equation}\label{eq:predictor-general-solution}
\begin{aligned}
\mathcal{P}
&=
Y_{\mathrm f}^{\mathrm d}\mathcal D^\dagger
+
\mathcal N_{\mathcal{P}}
\left(
    I_{(n_u+n_y)p+n_uf}
    -\mathcal D\mathcal D^\dagger
\right),\\
\mathcal N_{\mathcal{P}}
&\in
\mathbb R^{fn_y\times((n_u+n_y)p+n_uf)}.
\end{aligned}
\end{equation}
Indeed, the second term satisfies
\begin{equation}
\mathcal N_{\mathcal{P}}
\left(
    I_{(n_u+n_y)p+n_uf}
    -\mathcal D\mathcal D^\dagger
\right)
\mathcal D
=0.
\end{equation}
The unique minimum-Frobenius-norm minimizer is
\begin{equation}\label{eq:spc-predictor}
\mathcal{P}^\star
=
Y_{\mathrm f}^{\mathrm d}\mathcal D^\dagger.
\end{equation}

The predictor with fixed $\mathcal{P}$, commonly called the subspace predictor in subspace predictive control (SPC) \cite{favoreelSPCSubspacePredictive1999}, is defined as

\begin{equation}\label{eq:subspace-predictor}
y_k^{\mathrm{ind}} = \mathcal{P}z_k, \quad  z_k\coloneq\begin{bmatrix} u_{\mathrm{ini}} \\ y_{\mathrm{ini}} \\ u_k \end{bmatrix},
\end{equation}
with $u_{\mathrm{ini}}\coloneq u_{[k-p, k-1]} \in \mathbb{R}^{pn_u}$,  $ y_{\mathrm{ini}}\coloneq y_{[k-p, k-1]} \in \mathbb{R}^{pn_y}$, $u_k \coloneq u_{[k, k+f-1]} \in \mathbb{R}^{fn_u}$, and $y_k^{\mathrm{ind}} \coloneq y_{[k, k+f-1]} \in \mathbb{R}^{fn_y}$. The \emph{direct} predictor includes the coefficients that parameterize compatible trajectories through the offline data, as formalized by Willems' fundamental lemma \cite{willemsNotePersistencyExcitation2005}, and applied to constrained predictive control in \cite{coulsonDataEnabledPredictiveControl2019}. Since the present work starts from frequency-domain measurements, we next introduce the corresponding frequency-domain version of Willems' lemma, see \cite{meijerFrequencyDomainWillemsLemma2025a} for a complete overview.

\subsection{Frequency-domain Willems' fundamental lemma}

Let $\{\omega_m\}_{m\in\mathbb Z_{[0,M-1]}}$ denote the
equidistant frequency grid
\begin{equation}
\omega_m
\coloneqq
\frac{\pi m}{M},
\qquad
m\in\mathbb Z_{[0,M-1]}.
\end{equation} 

For a length-$2M$ sequence $v_{[0,2M-1]}$, the discrete Fourier transform (DFT) is given by
\begin{equation}\label{eq:dft}
V_m
\coloneqq
\sum_{k=0}^{2M-1}
v_k e^{-j\omega_m k},
\qquad
m\in\mathbb Z_{[0,M-1]}.
\end{equation}
Note that if $v_{[0,2M-1]} \in \mathbb{R}^{2M}$, then $V(-\omega_m) = V_m^*$ with $m \in \mathbb{Z}_{[1,M-1]}$, so $V_{[0,M-1]}$ and $V^*_{[1,M-1]}$ represent $2M-1$ frequencies and $V_0 \in \mathbb{R}$.

Applying the DFT to a length-$2M$ \emph{input-state-output} trajectory of
$\Sigma$ satisfying \eqref{eq:system} gives
\begin{align}
e^{j\omega_m}X_m
&=
AX_m+BU_m
+e^{j\omega_m}(x_0-x_{2M}),
\label{eq:dft-state-general}\\
Y_m
&=
CX_m+DU_m,
\label{eq:dft-output-general}
\end{align}
for all $m\in\mathbb Z_{[0,M-1]}$. In this work, the data are collected
in periodic steady state under multisine excitation. Hence, we assume
that
\begin{equation}
x_{2M}=x_0,
\end{equation}
such that \eqref{eq:dft-state-general}-\eqref{eq:dft-output-general}
reduce to
\begin{align}
e^{j\omega_m}X_m
&=AX_m+BU_m,
\label{eq:ss-spectrum-state}\\
Y_m
&=CX_m+DU_m,
\label{eq:ss-spectrum-output}
\end{align}
with $H_m \coloneqq H(e^{j\omega_m})
= C(e^{j\omega_m}I-A)^{-1}B+D$.

\begin{definition}[Steady-state input-output spectrum]
A pair of spectra
$(U_{[0,M-1]},Y_{[0,M-1]})$ is called a steady-state
input-output spectrum of $\Sigma$ if there exists a state spectrum
$X_{[0,M-1]}$ satisfying
\eqref{eq:ss-spectrum-state}-\eqref{eq:ss-spectrum-output} for all
$m\in\mathbb Z_{[0,M-1]}$. The corresponding triplet
$(U_{[0,M-1]},X_{[0,M-1]},Y_{[0,M-1]})$ is called a
steady-state input-state-output spectrum of $\Sigma$.
\end{definition}
\begin{remark}
Non-steady-state data can be handled by applying Willems' lemma to an augmented system that explicitly accounts for the term $(x_0-x_{2M})$, see \cite{meijerLeveragingNonSteadyStateFrequencyDomain2025a}.
\end{remark}

In the remainder, we restrict our attention to steady-state spectra. To connect steady-state spectra to finite trajectories, we collect the contribution of each spectral coefficient over a specified time interval. For integers $a<b$, define
\begin{equation}
W_{a,b}(z)
\coloneqq
\begin{bmatrix}
    z^a & z^{a+1} & \cdots & z^{b-1}
\end{bmatrix}^{\top},
\end{equation}
and
\begin{equation}
F_{a,b}(V_{[r,s]})
\coloneqq
\begin{bmatrix}
    W_{a,b}(e^{j\omega_r})\otimes V_r
    & \cdots &
    W_{a,b}(e^{j\omega_s})\otimes V_s
\end{bmatrix}.
\end{equation}
The complete conjugate-symmetric representation is
\begin{equation}
\Phi_{a,b}(V_{[0,M-1]})    \coloneqq
\begin{bmatrix}
    F_{a,b}(V_{[0,M-1]})
    &
    F_{a,b}^*(V_{[1,M-1]})
\end{bmatrix}.
\end{equation}

\begin{remark}
An equivalent real-valued representation is
\begin{equation}
\Phi_{a,b}^{\mathrm R}(V_{[0,M-1]})
\coloneqq
\begin{bmatrix}
    \operatorname{Re}F_{a,b}(V_{[0,M-1]})
    &
    \operatorname{Im}F_{a,b}(V_{[1,M-1]})
\end{bmatrix}.
\end{equation}
\end{remark}

\begin{lemma}[FD-WFL, {\cite[Lemma~1(ii)]{meijerFrequencyDomainWillemsLemma2025a}}]\label{lem:fd-wfl}
Let $(U_{[0,M-1]},Y_{[0,M-1]})$ be a steady-state
input-output spectrum of $\Sigma$, and suppose that
$U_{[0,M-1]}$ is persistently exciting of order $L+n_x$.
Then $(u_{[0,L-1]},y_{[0,L-1]})$ is a real length-$L$
input-output trajectory of $\Sigma$ if and only if there exists $G \coloneq \begin{bmatrix} G_0 & G_1^\top & (G_1^*)^\top\end{bmatrix}^\top \in \mathbb{C}^{2M-1}$ with $G_0 \in \mathbb{R}$ and $G_1\in\mathbb C^{M-1}$ such that
\begin{equation}\label{eq:FD-WFL}
\begin{bmatrix}
    u_{[0,L-1]}\\
    y_{[0,L-1]}
\end{bmatrix}
=
\begin{bmatrix}
    \Phi_{0,L}(U_{[0,M-1]})\\
    \Phi_{0,L}(Y_{[0,M-1]})
\end{bmatrix}G
\end{equation}
\end{lemma}
Since $W_{0,L}=W_L$, $F_{0,L}=F_L$, and
$\Phi_{0,L}=\Psi_L$, the standard notation is recovered by setting
$(a,b)=(0,L)$. The general indices will be used later for the input-output frequency-domain data-equation.

\subsection{Frequency-response functions and sample covariance}\label{sec:FRFprelim}
For ease of exposition, we consider a single-input single-output (SISO) setting with deterministic periodic excitation consisting of $P\in\mathbb Z_{\geq 2}$ independent repetitions. For each repetition $s\in\mathbb Z_{[1,P]}$, the measured time-domain output is

\begin{equation}
y_k^{[s]} = y_k + v_k^{[s]},
\end{equation}
where $y_k$ denotes the noise-free periodic steady-state output and $v_k^{[s]}$ the noise process modelled as
\begin{equation}
v_k^{[s]} = H_v(q)e_k^{[s]}.
\end{equation}
Here $H_v(q)$ is a monic, stable and linear filter and $e_k^{[s]}$ are zero-mean i.i.d. innovations independent across repetitions. For each repetition, the filter state is initialized independently in its stationary distribution. The resulting measured output spectrum satisfies
\begin{equation}
\hat{Y}_m^{[s]}
=
H_m U_m+ \mathcal{V}_m^{[s]},
\end{equation}
and with 
\begin{equation}\label{eq:frf-estimate}
\hat{H}_m^{[s]} = \frac{\hat{Y}_m^{[s]}}{U_m}, \quad \hat{H}_m = \frac{1}{P} \sum_{s=1}^{P} \hat{H}_m^{[s]},
\end{equation}
which denote the FRF estimate for repetition $s$ and the FRF estimate over all repetitions respectively. Thus, the complete FRF measurement over $M$ positive frequencies is denoted by $\hat{H}_{[0,M-1]}$. Define the difference $\Delta H_{[0,M-1]} \coloneq \hat{H}_{[0,M-1]} - H_{[0,M-1]}$, then the covariance matrix is defined as
\begin{equation}
C_H \coloneq \operatorname{Cov}(\hat H_{[0,M-1]})=\mathbb E\left[
\Delta H_{[0,M-1]} (\Delta H_{[0,M-1]})^H
\right],
\end{equation}
and by defining $\Delta \hat{H}_{[0,M-1]}^{[s]} \coloneq \hat H_{[0,M-1]}^{[s]}-\hat H_{[0,M-1]}$, then the sample covariance is defined as
\begin{equation}\label{eq:sample-covariance}
\hat{C}_H \coloneq \frac{1}{P(P-1)} \sum_{s=1}^{P}
\Delta \hat{H}_{[0,M-1]}^{[s]}
(\Delta \hat{H}_{[0,M-1]}^{[s]})^H,
\end{equation}
Thus, the transfer function estimate $\hat H_m$ and the associated sample covariance matrix $\hat{C}_H$ follow directly from
repeated periods without identifying a parametric noise model
\cite{pintelonSystemIdentificationFrequency2012}.

\subsection{Problem statement}
Given an estimated frequency response function $\hat{H}_{[0,M-1]}$ and its sample covariance $\hat{C}_H$ defined in \eqref{eq:sample-covariance}, we consider two problems. First, construct an \emph{indirect} FRF-based predictor $y_k^{\mathrm{ind}}$ and establish its equivalence with the \emph{direct} predictor $y_k^{\mathrm{dir}}$ introduced in \cite{meijerFrequencyDomainWillemsLemma2025a}, i.e., $y_k^{\mathrm{ind}}=y_k^{\mathrm{dir}}$ for all admissible initial trajectories and future inputs. Second, how to exploit the sample covariance $\hat C_H$ in both a FreePC and a FD-SPC formulation? To be specific, we aim minimize the expected squared error that originates from the use of $\hat{H}_{[0,M-1]}$ rather than the exact (unavailable) $H_{[0,M-1]}$ FRF measurement.

\section{FREQUENCY-DOMAIN SUBSPACE PREDICTIVE CONTROL}\label{sec:FD-dataeq}
Before we introduce frequency-domain subspace predictive control (FD-SPC) we first derive the frequency-domain data equation similar to the time-domain, see Sec.~\ref{sec:subspace-pred}.
\subsection{Frequency-domain input-output data equation}
Let $z_m\coloneqq e^{j\omega_m}$, then by
\eqref{eq:ss-spectrum-state} and \eqref{eq:ss-spectrum-output}, the
steady-state input-state-output spectrum
$(U_m,X_m,Y_m)$ which satisfies
$z_mX_m=AX_m+BU_m$ and
$Y_m=CX_m+DU_m$. Define the corresponding complex-valued
sequences as
\begin{equation}
u_k^{(m)}\coloneqq z_m^kU_m,\quad
x_k^{(m)}\coloneqq z_m^kX_m,\quad
y_k^{(m)}\coloneqq z_m^kY_m.
\end{equation}
These sequences form a trajectory of $\Sigma$, since
\begin{align}
x_{k+1}^{(m)}
&=z_m^k(z_mX_m)
=Ax_k^{(m)}+Bu_k^{(m)},\\
y_k^{(m)}
&=z_m^kY_m
=Cx_k^{(m)}+Du_k^{(m)}.
\end{align}
Their past and future trajectories are
\begin{align}
u_{[0,p-1]}^{(m)}
&=W_{0,p}(z_m)\otimes U_m,\\
y_{[0,p-1]}^{(m)}
&=W_{0,p}(z_m)\otimes Y_m,\\
u_{[p,p+f-1]}^{(m)}
&=W_{p,p+f}(z_m)\otimes U_m,\\
y_{[p,p+f-1]}^{(m)}
&=W_{p,p+f}(z_m)\otimes Y_m.
\end{align}
Applying \eqref{eq:io-data-1step} to this trajectory gives
\begin{equation}\label{eq:fd-io-component}
\begin{aligned}
W_{p,p+f}(z_m)\otimes Y_m
={}&
\mathcal O_f\mathcal K
\begin{bmatrix}
W_{0,p}(z_m)\otimes U_m\\
W_{0,p}(z_m)\otimes Y_m
\end{bmatrix}\\
&+\mathcal T_f
\bigl(W_{p,p+f}(z_m)\otimes U_m\bigr).
\end{aligned}
\end{equation}

Let $(U_{[0,M-1]},Y_{[0,M-1]})$ be a valid steady-state spectrum obtained
from the offline data. Applying \eqref{eq:fd-io-component} over all
admissible frequencies and their complex conjugates gives the following
\begin{equation}\label{eq:fd-io-data-equation}
\begin{aligned}
\Phi_{p,p+f}(Y_{[0,M-1]})
={}&
\mathcal O_f\mathcal K
\begin{bmatrix}
\Phi_{0,p}(U_{[0,M-1]})\\
\Phi_{0,p}(Y_{[0,M-1]})
\end{bmatrix}\\
&+
\mathcal T_f
\Phi_{p,p+f}(U_{[0,M-1]}),
\end{aligned}
\end{equation}
which we call the frequency-domain input-output data equation.
\begin{remark}
The columns of a block-Hankel matrix are shifted segments $v_{[i+a,i+b-1]}$, whereas the $m$-th column of $\Phi_{a,b}(V)$ is the restriction to $[a,b)$ of the $m$-th frequency of the complete time-domain sequence $v_{[0,N-1]}$, namely $W_{a,b}(e^{j\omega_m})\otimes V_m$. Hence, $\Phi_{a,b}(V)$ is not the columnwise DFT of a block-Hankel matrix. Even when both matrices span the same trajectory space, this equivalence applies to their images rather than their individual columns.
\end{remark}

\subsection{Indirect FD-SPC formulation}
From the frequency-domain input-output data equation we derive the indirect frequency-domain subspace predictor, for which the derivation is analogous to the time-domain variant as presented in Sec.~\ref{sec:subspace-pred}. We define the frequency-domain data matrix as
\begin{equation}
\bar{\mathcal{D}} = \begin{bmatrix}\Phi_{0,p}(U_{[0,M-1]})\\
\Phi_{0,p}(Y_{[0,M-1]}) \\ \Phi_{p,p+f}(U_{[0,M-1]})\end{bmatrix} \in \mathbb{C}^{((n_u+n_y)p+n_uf)\times(2M-1)},
\end{equation}
and then we can write \eqref{eq:fd-io-data-equation} compactly as
\begin{equation}
\Phi_{p,p+f}(Y_{[0,M-1]}) = \mathcal{P}\bar{\mathcal{D}},
\end{equation}
with $\mathcal{P}$ the same dimension as in \eqref{eq:P-dimension}, same general solution as in \eqref{eq:predictor-general-solution} and with the unique minimum-Frobenius-norm minimizer
\begin{equation}\label{eq:fd-spc-predictor}
\mathcal{P}^\star= \Phi_{p,p+f}(Y_{[0,M-1]}) \bar{\mathcal D}^\dagger.
\end{equation}
Therefore, the multistep predictor can be estimated directly from the input-output spectra $(U_{[0,M-1]}, Y_{[0,M-1]})$ without first identifying a state-space realization. 

By utilizing this subspace predictor in a receding horizon optimization problem, the future input sequence $u_k$ is subsequently optimized
online, whereas $\mathcal{P}^\star$ is fixed for all time, and $u_{\mathrm{ini}}$, and
$y_{\mathrm{ini}}$ are fixed at time $k$.
The frequency-domain subspace predictive control problem is defined as
\begin{equation}\label{eq:spc}
\begin{aligned}
\min_{u_k, y_k^{\mathrm{ind}}}\quad&
\|y_k^{\mathrm{ind}}- r^y_k\|_{\bar Q}^2+\|u_k - r^u_k\|_{\bar R}^2\\
\mathrm{s.t.}\quad&
y_k^{\mathrm{ind}}
=
\mathcal{P}^\star
z_k,\\
&
u_{i|k}\in\mathcal U \subseteq \mathbb{R}^{n_u},\quad
y_{i|k}^{\mathrm{ind}}\in\mathcal Y\subseteq \mathbb{R}^{n_y},
\quad i\in\mathbb Z_{[0,f-1]},
\end{aligned}
\end{equation}
where $\bar Q=I_f\otimes Q$, $\bar R=I_f\otimes R$, $Q\succeq0$, and $R\succ0$ with prediction horizon $f$. Only the first optimal input $u_{0|k}^\star$ is applied, after which the initial trajectory is updated and \eqref{eq:spc} is solved again.

\subsection{FreePC and noiseless equivalence to FD-SPC}

Let $\Phi_{\mathrm f}\coloneqq\Phi_{p,p+f}(Y_{[0,M-1]})$ and consider
a length-$(p+f)$ trajectory of $\Sigma$. Under the assumptions of Lemma~\ref{lem:fd-wfl}, its past and future trajectories can be written as
\begin{equation}\label{eq:direct-prediction}
\begin{bmatrix}
u_{\mathrm{ini}}\\ y_{\mathrm{ini}}\\ u_k\\ y_k^{\mathrm{dir}}
\end{bmatrix}
=
\begin{bmatrix}
\bar{\mathcal D}\\ \Phi_{\mathrm f}
\end{bmatrix}G_k.
\end{equation}
At time $k$, the measured $(u_{\mathrm{ini}},y_{\mathrm{ini}})$ is fixed, 
FreePC \cite{meijerFrequencyDomainWillemsLemma2025a} uses
\eqref{eq:direct-prediction} as its online prediction constraint and solves
\begin{equation}\label{eq:freepc}
\begin{aligned}
\min_{G_k,u_k,y_k^{\mathrm{dir}}}\quad&
\|y_k^{\mathrm{dir}}- r^y_k\|_{\bar Q}^2+\|u_k - r^u_k\|_{\bar R}^2\\
\mathrm{s.t.}\quad&
\bar{\mathcal D}G_k=z_k,\qquad y_k^{\mathrm{dir}}=\Phi_{\mathrm f}G_k,\\
&u_{i|k}\in\mathcal U \subseteq \mathbb{R}^{n_u},\;
y_{i|k}^{\mathrm{dir}}\in\mathcal Y\subseteq \mathbb{R}^{n_y},
\;i\in\mathbb Z_{[0,f-1]}.
\end{aligned}
\end{equation}
The first optimal input $u_{0|k}^\star$ is applied to $\Sigma$. For a given $z_k$, the constraint $\bar{\mathcal D}G_k=z_k$ may admit several coefficients, namely the general solution
\begin{equation}\label{eq:generalsol-freepc}
G_k=\bar{\mathcal D}^{\dagger}z_k+\left(I_{2M-1}-\bar{\mathcal D}^{\dagger}\bar{\mathcal D}\right)w_k.
\end{equation}
The following proposition shows that, with noiseless data, all of them yield the FD-SPC prediction $\mathcal P^\star z_k$ by following a similar reasoning as in \cite{fiedlerRelationshipDataenabledPredictive2021}, but by using the singular value decomposition.

\begin{proposition}[Equivalence of FD-SPC and FreePC]\label{lem:equivalence}
Consider the LTI system $\Sigma$ in \eqref{eq:system}, and let
$p\geq\ell_\Sigma$ and $f\geq1$, and let $\{H_m\}_{m=0}^{M-1}$ be exact FRF samples of $\Sigma$ on the frequency grid. Let $(U_{[0,M-1]},Y_{[0,M-1]})$ be an exact steady-state input-output spectrum of $\Sigma$, with $U_{[0,M-1]}$ persistently exciting of order $p+f+n_x$. Then, for every initial trajectory $(u_{\mathrm{ini}},y_{\mathrm{ini}})$ compatible with $\Sigma$ and every future input $u_k$, FD-SPC and FreePC yield the same
prediction, $y_k^{\mathrm{ind}}=y_k^{\mathrm{dir}}$. 
\end{proposition}
\begin{proof}
Let  $\rho\coloneqq\rank(\bar{\mathcal D})$ and consider the compact SVD $\bar{\mathcal D}=U_\rho\Sigma_\rho V_\rho^H$, completed by $U=[\,U_\rho\ U_0\,]$ and $V=[\,V_\rho\ V_0\,]$, so that $\bar{\mathcal D}^\dagger=V_\rho\Sigma_\rho^{-1}U_\rho^H$, $I-\bar{\mathcal D}\bar{\mathcal D}^\dagger=U_0U_0^H$, and $I-\bar{\mathcal D}^\dagger\bar{\mathcal D}=V_0V_0^H$. By Lemma~\ref{lem:fd-wfl}, $z_k\in\im(\bar{\mathcal D})$, and hence $U_0^Hz_k=0$. Therefore, every indirect predictor $\mathcal{P}=\Phi_{\mathrm f}\bar{\mathcal D}^\dagger+\mathcal N_{\mathcal{P}}U_0U_0^H$ satisfies $\mathcal{P}z_k=\Phi_{\mathrm f}\bar{\mathcal D}^\dagger z_k$. Every solution of $\bar{\mathcal D}G=z_k$ is of the form $G=\bar{\mathcal D}^\dagger z_k+V_0\eta$. Since $\bar{\mathcal D}V_0=0$, the frequency-domain data equation \eqref{eq:fd-io-data-equation} gives $\Phi_{\mathrm f}V_0=0$. It follows that $y_k^{\mathrm{ind}}=\mathcal{P}z_k=\Phi_{\mathrm f}\bar{\mathcal D}^\dagger z_k=\Phi_{\mathrm f}G= y_k^{\mathrm{dir}}$. 
\end{proof}

Therefore, under identical costs and constraints, both yield the same prediction and hence the same optimal future inputs. In the case of noisy data, this result cannot directly be used, however, there are established time-domain equivalences by utilizing $\Pi$-regularization or by using instrumental variables (IVs), see \cite{dorflerBridgingDirectIndirect2023} and \cite{wingerdenDataenabledPredictiveControl2022}.

\section{SAMPLE-COVARIANCE REGULARIZATION}
In Sec.~\ref{sec:FD-dataeq}, the data matrices are constructed from an exact input-output spectrum. We now construct the same matrices from repeated FRF estimates and use the variation across repetitions to regularize FD-SPC and FreePC.
\subsection{Offline and online data uncertainty}
Let $\Omega\subseteq\mathbb Z_{[0,M-1]}$ denote the set of selected
frequency indices. Following
\cite[Remark~1]{meijerFrequencyDomainWillemsLemma2025a}, we construct
the input-output spectrum using $U_m=1$ for $m\in\Omega$. At these
frequencies, the exact output spectrum, its estimate from
repetition $s$, and its average over $P$ repetitions are
\begin{equation}
Y_m=H_m,\qquad
\hat Y_m^{[s]}=\hat H_m^{[s]},\qquad
\hat Y_m=\hat H_m,
\end{equation}
respectively. We set all constructed spectra to zero outside $\Omega$. 

Under the measurement assumptions of Sec.~\ref{sec:FRFprelim}, $\mathbb E[\hat Y_m]=Y_m$. Using the averaged output spectrum gives the offline data matrices
\begin{equation}
\begin{aligned}
\hat{\mathcal D}
&\coloneqq
\begin{bmatrix}
\Phi_{0,p}(U_{[0,M-1]})\\
\Phi_{0,p}(\hat Y_{[0,M-1]})\\
\Phi_{p,p+f}(U_{[0,M-1]})
\end{bmatrix},\\
\hat\Phi_{\mathrm f}
&\coloneqq \Phi_{p,p+f}(\hat Y_{[0,M-1]}).
\end{aligned}
\end{equation}
Since the constructed input spectrum is fixed, only the output
blocks depend on the FRF estimation error. With
$\Delta Y_{[0,M-1]}\coloneqq\hat Y_{[0,M-1]}-Y_{[0,M-1]}$,
the corresponding errors are
\begin{equation}
\begin{aligned}
\Delta\mathcal D
&\coloneqq \hat{\mathcal D}-\bar{\mathcal D}
=\begin{bmatrix}
0\\
\Phi_{0,p}(\Delta Y_{[0,M-1]})\\
0
\end{bmatrix},\\
\Delta\Phi
&\coloneqq \hat\Phi_{\mathrm f}-\Phi_{\mathrm f}
=\Phi_{p,p+f}(\Delta Y_{[0,M-1]}).
\end{aligned}
\end{equation}
Both errors have zero mean and depend on the same output-spectrum error.

To construct the regularizers, we use the observed variation across
repetitions. Let $\hat{\mathcal D}^{[s]}$ denote the data matrix
constructed from $\hat Y_{[0,M-1]}^{[s]}$. Its mean over the $P$
repetitions and its deviation from that mean satisfy
\begin{equation}
\hat{\mathcal D}
=\frac{1}{P}\sum_{s=1}^{P}\hat{\mathcal D}^{[s]},
\qquad
\Delta\hat{\mathcal D}^{[s]}
\coloneqq\hat{\mathcal D}^{[s]}-\hat{\mathcal D}.
\end{equation}
For $m\in\Omega$, we have
$\hat Y_m^{[s]}-\hat Y_m=\hat H_m^{[s]}-\hat H_m$.
Thus, the deviations $\Delta\hat{\mathcal D}^{[s]}$ used in both
regularizers are constructed from the same centered FRF estimates
as the sample covariance $\hat C_H$ in \eqref{eq:sample-covariance}.

Online, the inputs are assumed to be exact and the measured initial output is denoted by $\hat y_{\mathrm{ini}}$. The measured data vector and its error are therefore
\begin{equation}
\hat z_k\coloneqq
\begin{bmatrix}
u_{\mathrm{ini}}\\
\hat y_{\mathrm{ini}}\\
u_k
\end{bmatrix},
\qquad
\Delta z_k\coloneqq\hat z_k-z_k.
\end{equation}
We assume that the initial-output error
$\hat y_{\mathrm{ini}}-y_{\mathrm{ini}}$ has zero mean and is
independent of the offline output-spectrum error $\Delta Y_{[0,M-1]}$.
In the following derivations, expectations are taken for fixed
$\mathcal P$ in FD-SPC and fixed $G_k$ in FreePC.

\subsection{FD-SPC sample-variance regularization}\label{sec:ind-var}
In FD-SPC, uncertainty in the FRF estimates affects both matrices
in the predictor fit. We compute the predictor from the regularized
least-squares problem
\begin{equation}\label{eq:fdreg-prob}
\min_{\mathcal P}\;
\|\hat\Phi_{\mathrm f}-\mathcal P\hat{\mathcal D}\|_{\mathrm F}^2
+\operatorname{tr}\!\left(
  \mathcal P R_{\mathcal P}\mathcal P^H
\right),
\end{equation}
where the weight is computed from the deviations across repetitions as
\begin{equation}
R_{\mathcal P}\coloneqq
\frac{1}{P(P-1)}
\sum_{s=1}^{P}
\Delta\hat{\mathcal D}^{[s]}
(\Delta\hat{\mathcal D}^{[s]})^H \in \mathbb{C}^{n_z\times n_z},
\end{equation}
where $n_z:=(n_u+n_y)p+n_uf$. The quadratic term penalizes predictor coefficients that amplify
errors in the past-output data. To motivate this penalty, consider the expected fitting residual
for fixed $\mathcal P$. Substituting
$\hat{\mathcal D}=\bar{\mathcal D}+\Delta\mathcal D$ and
$\hat\Phi_{\mathrm f}=\Phi_{\mathrm f}+\Delta\Phi$ and using
$\mathbb E[\Delta\mathcal D]=0$ and $\mathbb E[\Delta\Phi]=0$ gives
\begin{equation}
\begin{aligned}
&\mathbb E\!\left[
  \|\hat\Phi_{\mathrm f}-\mathcal P\hat{\mathcal D}\|_{\mathrm F}^2
\right]\\
&\quad=
\|\Phi_{\mathrm f}-\mathcal P\bar{\mathcal D}\|_{\mathrm F}^2
+\mathbb E[\|\Delta\Phi\|_{\mathrm F}^2]\\
&\qquad+
\operatorname{tr}\!\left(
  \mathcal P\,\mathbb E[\Delta\mathcal D\Delta\mathcal D^H]\,
  \mathcal P^H
\right)
-2\operatorname{Re}\operatorname{tr}
  (\mathcal P C_{\mathcal D,\Phi}),
\end{aligned}
\end{equation}
where $C_{\mathcal D,\Phi}\coloneqq
\mathbb E[\Delta\mathcal D\Delta\Phi^H]$.
The term $\mathbb E[\|\Delta\Phi\|_{\mathrm F}^2]$ is independent
of $\mathcal P$. The cross term can be nonzero because the past
and future output blocks depend on the same FRF estimates.
The penalty in \eqref{eq:fdreg-prob} retains the quadratic
contribution, with $R_{\mathcal P}$ estimating
$\mathbb E[\Delta\mathcal D\Delta\mathcal D^H]$.

The minimum-Frobenius-norm solution of \eqref{eq:fdreg-prob} is
\begin{equation}
\mathcal P^\star
=
\hat\Phi_{\mathrm f}\hat{\mathcal D}^H
\left(
  \hat{\mathcal D}\hat{\mathcal D}^H+R_{\mathcal P}
\right)^\dagger.
\end{equation}
This predictor is computed offline. At time $k$, FD-SPC uses
$y_k^{\mathrm{ind}}=\mathcal P^\star\hat z_k$ in \eqref{eq:spc}.

\subsection{FreePC sample-covariance regularization}\label{sec:dir-var}
For FreePC, we include the slack $\sigma_k\in\mathbb R^{pn_y}$ to allow a mismatch
between the measured and fitted initial outputs, with penalty
$\lambda_\sigma\|\sigma_k\|_2^2$. The term $\lambda_G\|G_k\|_2^2$
discourages large coefficients. To account for uncertainty
in the FRF estimates, we add a covariance-weighted penalty on $G_k$,
resulting into the covariance regularized FreePC problem 
\begin{equation}\label{eq:weighted-freepc}
\begin{aligned}
\min_{\substack{u_k,y_k^{\mathrm{dir}},\\G_k,\sigma_k}}\quad&
J_k^{\mathrm{dir}}+\|R_G^{1/2}G_k\|_2^2
+\lambda_G\|G_k\|_2^2+\lambda_\sigma\|\sigma_k\|_2^2\\
\mathrm{s.t.}\quad&
\hat{\mathcal D}G_k
=\hat z_k+\begin{bmatrix}0\\ \sigma_k\\ 0\end{bmatrix},\\
&y_k^{\mathrm{dir}}=\hat\Phi_{\mathrm f}G_k,\\
&u_{i|k}\in\mathcal U,\quad y_{i|k}^{\mathrm{dir}}\in\mathcal Y,
\quad i\in\mathbb Z_{[0,f-1]},
\end{aligned}
\end{equation}
where $\lambda_G,\lambda_\sigma>0$ and $G_k$ has the
conjugate-symmetric form in Lemma~\ref{lem:fd-wfl}.
The covariance weight is computed from the deviations across
repetitions as
\begin{equation}
R_G\coloneqq
\frac{1}{P(P-1)}
\sum_{s=1}^{P}
(\Delta\hat{\mathcal D}^{[s]})^H
\Delta\hat{\mathcal D}^{[s]} \in \mathbb{C}^{(2M-1)\times (2M-1)}
\end{equation}
The quadratic term $G_k^H R_G G_k$ penalizes combinations of trajectory coefficients that amplify errors in the past-output data. 

To motivate this penalty, consider the expected squared residual in the trajectory constraint for fixed $G_k$ and $z_k$.
Substituting $\hat z_k=z_k+\Delta z_k$ and
$\hat{\mathcal D}=\bar{\mathcal D}+\Delta\mathcal D$, and using
the independence of the zero-mean errors $\Delta z_k$ and
$\Delta\mathcal D$, gives
\begin{equation}
\begin{aligned}
&\mathbb E\!\left[
  \|\hat z_k-\hat{\mathcal D}G_k\|_2^2
\right]\\
&\quad=
\|z_k-\bar{\mathcal D}G_k\|_2^2
+\mathbb E[\|\Delta z_k\|_2^2]\\
&\qquad+
G_k^H\mathbb E[\Delta\mathcal D^H\Delta\mathcal D]G_k.
\end{aligned}
\end{equation}
The online error term $\mathbb E[\|\Delta z_k\|_2^2]$ is independent
of $G_k$. The final term equals
$\mathbb E[\|\Delta\mathcal D G_k\|_2^2]$, the expected squared
contribution of the offline data error. Estimating
$\mathbb E[\Delta\mathcal D^H\Delta\mathcal D]$ by $R_G$ gives
the covariance penalty in \eqref{eq:weighted-freepc}.

\section{NUMERICAL CASE STUDY}\label{sec:numerical-example}
This section illustrates the proposed methods in Sec.~\ref{sec:dir-var} and Sec.~\ref{sec:ind-var}. In particular, we investigate the effect of the number of measured periods and the noise spectrum on the resulting control cost within one horizon length. 

\subsection{System, parameters, and performance metric}
Consider the benchmark system given in \cite{landauFlexibleTransmissionSystem1995}, which is given by the following transfer function 
\begin{equation}\label{eq:benchmark}
G(z) = \frac{0.2826z + 0.5067}{z^4 -1.4183z^3+1.5894z^2-1.3161z+0.8864}
\end{equation}
The system is excited by a random-phase multisine with $u_{\mathrm{rms}}=10$ at $M=20$ equidistant nonnegative frequencies, with period length $N=40$. The frequency response is estimated by averaging over $P$ measured periods according to \eqref{eq:frf-estimate}, with $P\in\{3,5,10,20\}$ in the control experiments.

\begin{figure}[b!]
\centering
\includegraphics[width=\columnwidth]{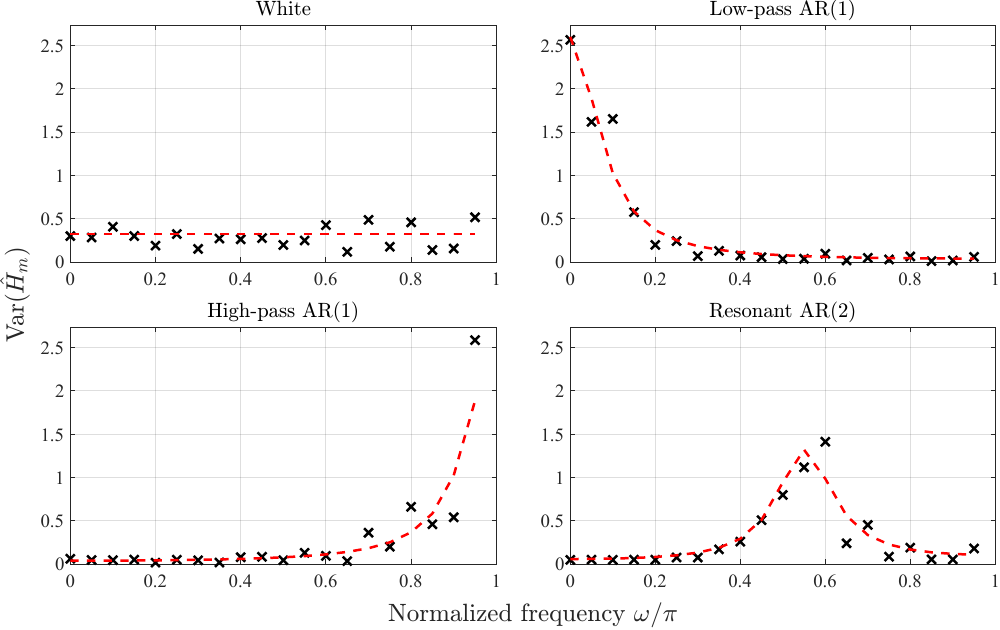}
\caption{Sample variance estimates (\frfSampleMarker) and exact variances (\frfVarianceLine) of the mean FRF estimate for $P=3$ under the corresponding noise filter $H_v$.}
\label{fig:samplevar}
\end{figure}

We consider white (W), low-pass (LP), high-pass (HP), and resonant (R)
noise, generated as $v_k=c_jH_j(q)e_k$, where $e_k\sim\mathcal{N}(0,1)$
are i.i.d. and
\begin{equation}
\begin{aligned}
H_{\mathrm W}(z)&=1,\\
H_{\mathrm{LP}}(z)&=(1-az^{-1})^{-1},\\
H_{\mathrm{HP}}(z)&=(1+az^{-1})^{-1},\\
H_{\mathrm R}(z)&=
(1-2a\cos(\omega_0)z^{-1}+a^2z^{-2})^{-1}.
\end{aligned}
\end{equation}
We use $a=0.8$, $\omega_0=0.55\pi$, and stationary initialization, choosing $c_j$ such that $\sigma_v=10$ and hence $\operatorname{Var}(v_k)=100$ for every noise model. The sample variance estimates are shown in Fig.~\ref{fig:samplevar} for $P=3$ which illustrates the finite-sample effects alongside the true variance which is unavailable from finite data.

Performance is evaluated over a single horizon of $f$ time steps. The input sequence $u$ is optimized once and applied to the true system without reoptimization. The resulting tracking cost is
\begin{equation}
J_{\mathrm{sys}}
=
\left\| y_{\mathrm{sys}}-y_{\mathrm{ref}}\right\|_{\bar Q}^{2}
+
\left\| u-u_{\mathrm{ref}}\right\|_{\bar R}^{2},
\end{equation}
where $y_{\mathrm{sys}}$ is the system output over that horizon. Regularization and slack penalties are excluded from this performance metric. The system starts from rest with zero input-output history and tracks a constant output reference of one with the corresponding steady-state input reference. We impose no bounds on future inputs or outputs. The boxplots shown in Fig.~\ref{fig:indbox} and Fig.~\ref{fig:dirbox} summarize 1000 Monte Carlo trials where all compared methods use the same excitation, initial trajectory, and reference. The past and future horizons are $p=10$ and $f=20$, respectively, with cost weights $\bar Q=I_f$ and $\bar R=10^{-1}I_f$. The oracle uses the exact benchmark model \eqref{eq:benchmark} to minimize the tracking cost over the same horizon with identical conditions. 

\subsection{FD-SPC sample-covariance regularization}
\begin{figure}[t]
\centering
\includegraphics[width=\columnwidth]{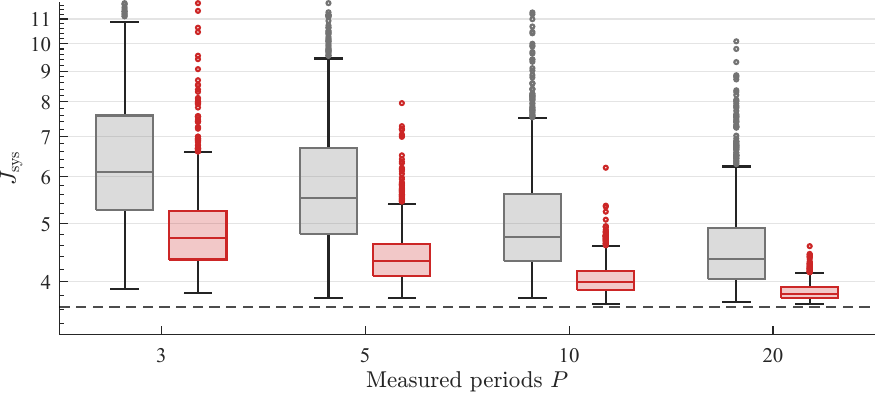}
\caption{Realized FD-SPC cost under resonant measurement noise over 1000 Monte Carlo trials: baseline (\controlBox{controlGray}), regularized (\controlBox{predictionRed}), and model oracle (\controlOracleLine). Boxes show medians and quartiles, circles mark outliers.}
\label{fig:indbox}
\end{figure}

To summarize the realized control cost $J_{\mathrm{sys}}$ for FD-SPC all 1000 Monte Carlo trials a boxplot is used and depicted in Fig.~\ref{fig:indbox} for the following set measured periods $P\in\{3,5,10,20\}$. For all trials the resonant noise filter $H_R(z)$ is used, for other colored noises the reader is referred to Sec.~\ref{sec:colored-noises}. The boxplot illustrates under these circumstances that including the regularization term $R_\mathcal{P}$ reduces the median realized control cost as well as its lower and upper quartiles. Furthermore, increasing the amount of periods $P$ consistently reduces the realized median control cost as well as its variability. The regularization is particularly useful in the case of few data, in this case, using our method for $P=3$ achieves
a comparable median cost compared to the baseline implementation for $P=10$, which can benefit systems where measurement time is costly.  

\subsection{FreePC sample-covariance regularization}
\begin{figure}[t]
\centering
\includegraphics[width=\columnwidth]{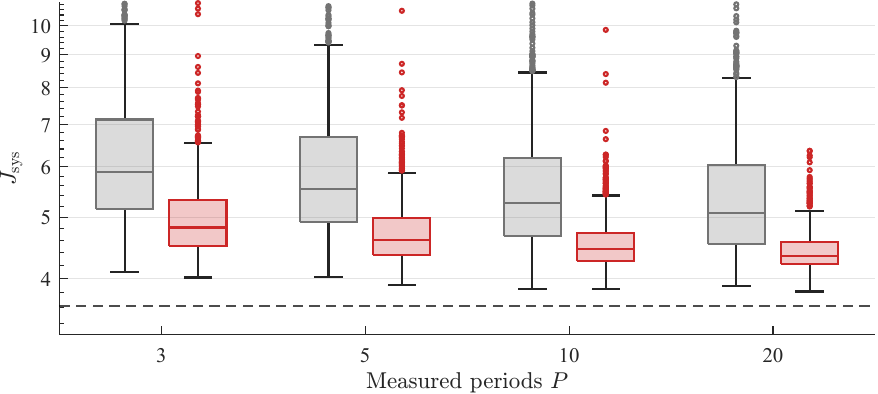}
\caption{Realized FreePC cost under resonant measurement noise over 1000 Monte Carlo trials: baseline (\controlBox{controlGray}), regularized (\controlBox{predictionRed}), and model oracle (\controlOracleLine). Boxes show medians and quartiles, circles mark outliers.}
\label{fig:dirbox}
\end{figure}

The FreePC formulation \eqref{eq:weighted-freepc} includes the parameters $\lambda_G$ and $\lambda_\sigma$, which generally require tuning. In this case study, we fix $(\lambda_G,\lambda_\sigma)=(10^{-8},1)$ without optimizing them. Our aim is to assess the benefit of the proposed regularizer within each formulation, rather than to compare the performance of direct and indirect control. Similarly, to FD-SPC, the resonant filter $H_R(z)$ is used, and the baseline FreePC is compared to the regularization term $R_G$. To obtain a simple frequency-wise uncertainty penalty, we approximate $R_G$ by $\operatorname{diag}(R_G)$, where $\operatorname{diag}(R_G)$ denotes the diagonal matrix containing its diagonal entries. This neglects correlations between frequencies and yields the penalty $\|\operatorname{diag}(R_G)^{1/2}G_k\|_2^2$. Thus, coefficients in $G_k$
associated with more uncertain FRF estimates are penalized more strongly. The realized cost $J_{\mathrm{sys}}$ in FreePC for 1000 Monte Carlo trials is depicted in Fig.~\ref{fig:dirbox} for the following set measured periods $P\in\{3,5,10,20\}$. The regularization consistently reduces the median realized control cost as well as its lower and upper quartiles for increasing number of periods $P$. Interestingly, utilizing the regularization $\|\operatorname{diag}(R_G)^{1/2}G_k\|_2^2$ for as low as $P=3$ periods can outperform the baseline realized control cost for $P=20$ periods, thus regularizing can benefit systems where measurement time is costly.

\begin{figure}[b]
\centering
\includegraphics[width=\columnwidth]{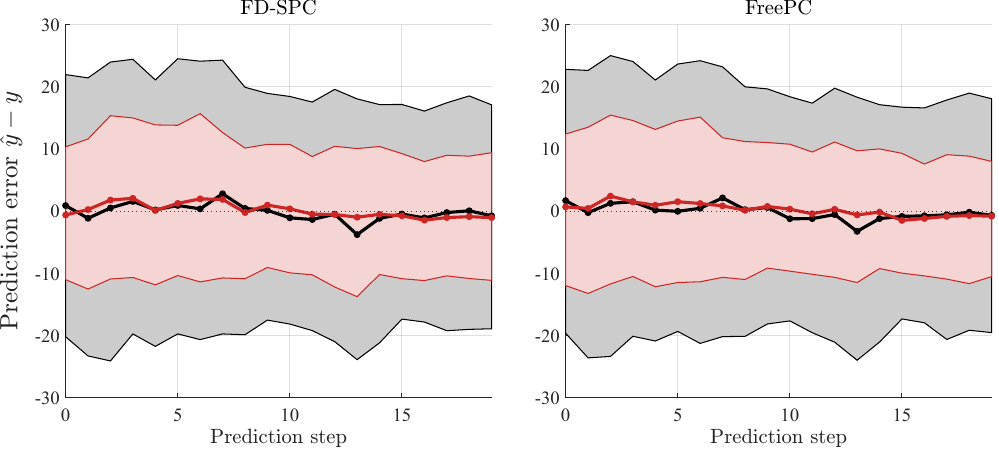}
\caption{Prediction errors $\hat y-y$ for the indirect (left) and direct (right) predictors under low-pass measurement noise with $P=10$. Baseline (\predictionLegend{black}{black!20}) and FRF-regularized (\predictionLegend{predictionRed}{predictionRed!20}) predictions are summarized over 1000 Monte Carlo trials. Dots show the median prediction error and the shaded band spans the first and third quartiles at each prediction step.}
\label{fig:prediction-errors}
\end{figure}

To examine prediction accuracy independently of control optimization, we evaluate the indirect and direct predictors on the same fixed validation trajectory. Across 1000 Monte Carlo trials, only the noise realization in the identification data varies. Each trial uses $P=10$ periods and the low-pass noise filter $H_{\mathrm{LP}}(z)$. Fig.~\ref{fig:prediction-errors} shows the median prediction error and the band between the first and third quartiles at each prediction step, with and without FRF regularization.

\subsection{Colored noise case study}\label{sec:colored-noises}
The median realized control cost $J_{\mathrm{sys}}$ for FreePC and FD-SPC under white (W), low-pass (LP), high-pass (HP), and resonant (R) noise is shown in Tab.~\ref{tab:table1}. For each number of measured periods $P\in\{3,5,10,20\}$, the regularized cost appears in bold, followed by the baseline cost in parentheses. The proposed methods consistently outperform the baseline implementations for all colored-noise cases, and are lower bounded by the model-oracle cost which is 3.629.  

\begin{table}[t]
\centering
\caption{Median realized cost $J_{\mathrm{sys}}$ over 1000 Monte Carlo trials. Bold values are regularized costs, parentheses give the corresponding baseline costs}
\label{tab:table1}
\begingroup
\small
\setlength{\tabcolsep}{2pt}
\begin{tabular*}{\columnwidth}{@{\extracolsep{\fill}}lllll@{}}
\hline
$P$ & W & LP & HP & R\\
\hline
\multicolumn{5}{c}{FreePC}\\
3  & \textbf{5.99} (7.54) & \textbf{10.96} (12.97) & \textbf{4.75} (5.55) & \textbf{4.82} (5.89)\\
5  & \textbf{5.37} (6.77) & \textbf{8.70} (10.84)  & \textbf{4.57} (5.36) & \textbf{4.60} (5.53)\\
10 & \textbf{4.93} (6.09) & \textbf{6.59} (8.46)   & \textbf{4.45} (5.22) & \textbf{4.45} (5.27)\\
20 & \textbf{4.65} (5.71) & \textbf{5.46} (7.39)   & \textbf{4.37} (5.09) & \textbf{4.35} (5.07)\\
\hline
\multicolumn{5}{c}{FD-SPC}\\
3  & \textbf{6.27} (8.15) & \textbf{9.77} (14.49) & \textbf{4.23} (5.37) & \textbf{4.74} (6.10)\\
5  & \textbf{5.38} (7.23) & \textbf{7.87} (12.18) & \textbf{4.00} (4.92) & \textbf{4.33} (5.52)\\
10 & \textbf{4.69} (6.06) & \textbf{6.19} (9.16)  & \textbf{3.83} (4.52) & \textbf{4.00} (4.75)\\
20 & \textbf{4.19} (5.24) & \textbf{5.08} (7.23)  & \textbf{3.74} (4.20) & \textbf{3.82} (4.36)\\
\hline
\end{tabular*}
\endgroup
\end{table}

\section{CONCLUSIONS}
In this paper, we introduced two methods to exploit sample variance information from FRFs with repeated excitation signals in data-driven predictive control. We showed that exploiting this sample variance gives enhanced performance on the realized cost for a specific set of data lengths, and we showed equally consistent results under colored measurement noise. We introduced a new subspace-predictor that can be estimated directly from FRF data and showed equivalence to FreePC in the noiseless case.

\setcounter{subsection}{0}
\renewcommand{\thesubsection}{\Alph{subsection}}

\section*{ACKNOWLEDGMENT}
The authors used Codex to assist with implementing the numerical example in Section~\ref{sec:numerical-example}. The authors edited the content and take responsibility for the contents of the paper.
\bibliographystyle{IEEEtran}
\bibliography{references}

\end{document}